\documentclass{webofc}
\usepackage[varg]{txfonts}   
\usepackage{tikz}

\newcommand{\pp}{\ensuremath{\text{p\kern-0.05em p}}}

\newcommand{\PbPb}{\ensuremath{\mbox{Pb--Pb}}}
\newcommand{\AuAu}{\ensuremath{\mbox{Au--Au}}}
\newcommand{\sqrts}{\ensuremath{\sqrt{s_{\text{NN}}}}}

\newcommand{\GeVc}{\ensuremath{\text{GeV}\kern-0.05em/\kern-0.02em c}}

\newcommand{\pT}{\ensuremath{p_{\text{T}}}}

\newcommand{\Raa}{\ensuremath{R_{\text{AA}}}}

\newcommand{\qhat}{\ensuremath{\hat{q}}}

\begin{document}
\title{Measuring jet quenching with a Bayesian inference analysis of hadron and jet data by JETSCAPE}
%
%

\author{\firstname{Raymond} \lastname{Ehlers}\inst{1,2}\fnsep\thanks{\email{raymond.ehlers@cern.ch}} for the JETSCAPE Collaboration
}

\institute{University of California, Berkeley 
\and
           Lawrence Berkeley National Laboratory
          }

\abstract{%
The JETSCAPE Collaboration reports the first multi-messenger study of the QGP jet transport parameter $\hat{q}$ using Bayesian inference, incorporating all available hadron and jet inclusive yield and jet substructure data from RHIC and the LHC. The theoretical model utilizes virtuality-dependent in-medium partonic energy loss coupled to a detailed dynamical model of QGP evolution. Tension is observed when constraining $\hat{q}$ for different kinematic cuts of the inclusive hadron data. The addition of substructure data is shown to improve the constraint on $\hat{q}$, without inducing tension with the constraint due to inclusive observables. These studies provide new insight into the mechanisms of jet interactions in matter, and point to next steps in the field for comprehensive understanding of jet quenching as a probe of the QGP.
}
\maketitle

\section{Introduction}\label{sec:introduction}

Jet quenching measurements at RHIC and the LHC provide a wealth of information about the quark-gluon plasma. However, different model approaches incorporating jet quenching, which are based on different formulations of the underlying physics, can describe the same inclusive yield suppression data equally well. Discrimination of these different physical pictures requires more rigorous and systematic comparisons of multi-messenger data. Bayesian inference provides the suitable framework for such a program. In heavy-ion physics, Bayesian inference has been applied successfully in both the soft and hard sectors~\cite{Bernhard:2016tnd,Novak:2013bqa,JETSCAPE:2020mzn,JETSCAPE:2021ehl}. Its application combining high-$\pT{}$ hadron and jet measurements is less well developed, and is discussed here.

The analysis utilizes a multi-stage approach with MATTER+LBT implemented in the JETSCAPE simulation framework~\cite{Cao:2017qpx,Cao:2017hhk,JETSCAPE:2017eso}.
Partons are propagated in a 2+1D medium simulated via relativistic viscous hydrodynamics which is calibrated to soft sector observables~\cite{Bernhard:2019bmu}.
Jet energy loss is parametrized based on a hard thermal loop calculation of the jet transport coefficient $\qhat{}$, modulated by a virtuality dependent term which reduces energy loss for highly virtuality partons~\cite{,JETSCAPE:2022jer}.
The full form of the parametrization is provided in Ref.~\cite{Ehlers:2022ulm}.
The calculations required 10 million core-hours on high performance computing facilities provided by XSEDE~\cite{Towns:2014qtb}.

\section{Bayesian inference with inclusive hadron and jet $\Raa{}$}\label{sec:hadronJetRaa}

The initial JETSCAPE proof-of-principle analysis to constrain $\qhat{}$ only utilized inclusive charged hadron $\Raa{}$~\cite{JETSCAPE:2021ehl}. The next step is to add the jet $\Raa{}$.
The analysis includes all applicable data which was published at time of the calculations.
We utilize inclusive charged hadron $\Raa{}$ and jet $\Raa{}$ from ALICE, ATLAS, CMS, PHENIX, and STAR measured in $\AuAu{}$ collisions at 200 GeV at RHIC and $\PbPb{}$ collisions at 2.76 TeV and 5.02 TeV at the LHC~\cite{ALICE:2015dtd, ALICE:2018mdl, ALICE:2018vuu, ATLAS:2015qmb, CMS:2012aa, ALICE:2015mjv, ATLAS:2014ipv, CMS:2016uxf, ALICE:2018vuu, ALICE:2019hno, CMS:2016xef, ALICE:2019qyj, ATLAS:2018gwx, CMS:2021vui, PHENIX:2012jha, STAR:2003fka, STAR:2020xiv}.
All data are either central or semi-central, i.e. within 0--50\% centrality.

\begin{figure}[t]
    \centering
    \begin{tikzpicture}
        \node[anchor=south west, inner sep=0] (image) at (0,0) {\includegraphics[height=5.5cm]{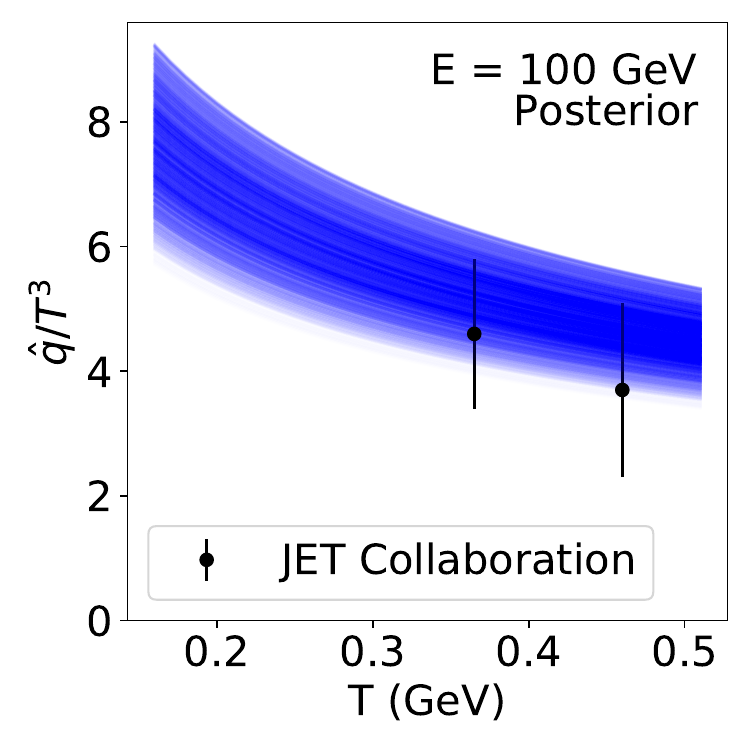}};
        \begin{scope}[x={(image.south east)},y={(image.north west)}]
            \node[anchor=south west, font=\sffamily\footnotesize] at (0.18, 0.30) {JETSCAPE Preliminary};
        \end{scope}
    \end{tikzpicture}
    \begin{tikzpicture}
        \node[anchor=south west, inner sep=0] (image) at (0,0) {\includegraphics[height=5.5cm]{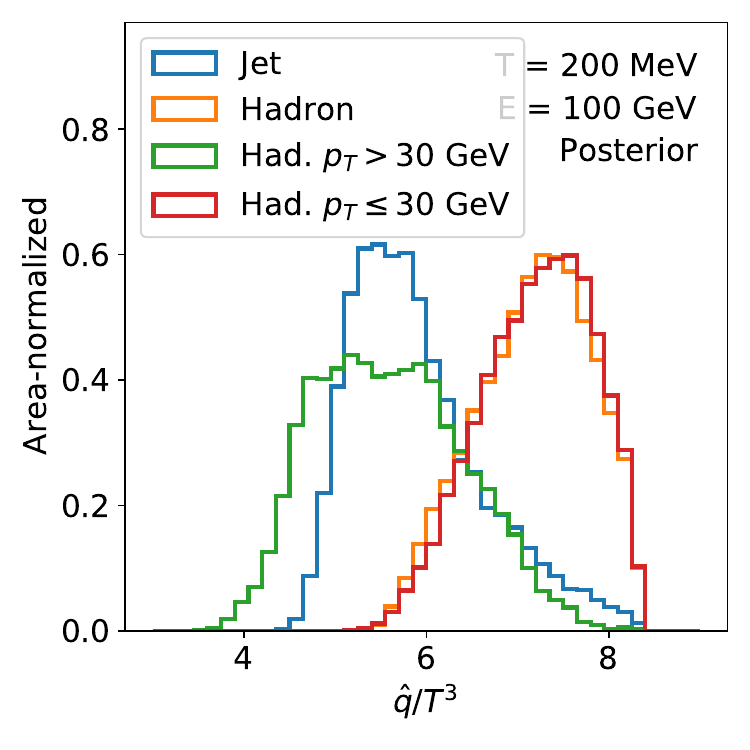}};
        \begin{scope}[x={(image.south east)},y={(image.north west)}]
            \node[anchor=north east, font=\sffamily\scriptsize, align=center, execute at begin node=\setlength{\baselineskip}{6pt}] at (0.45, 0.675) {JETSCAPE\\Preliminary\par};
        \end{scope}
    \end{tikzpicture}
    \caption{Medium $T$ dependent posterior distribution of $\qhat{}$ for the inclusive hadron and jet $\Raa{}$ analysis (left), and projections of the $\qhat{}$ posterior distribution at fixed medium $T$ and parton $E$ (right) for separate classes of data (see text). 
    }
    \label{fig:hadronJetRaaPosteriors}
\end{figure}

The Bayesian inference analysis follows Ref.~\cite{JETSCAPE:2021ehl}, including similar treatment of experimental uncertainties.
The medium temperature dependence of the $\qhat{}$ posterior distribution is shown on the left side of Fig.~\ref{fig:hadronJetRaaPosteriors}, where the blue band is generated by sampling within the 90\% confidence interval of the posterior.
This demonstrates that it is possible to extract a consistent $\qhat{}$ distribution when including all available inclusive hadron and jet $\Raa{}$ measurements.
The distribution is consistent within uncertainties with previous extractions by the JET collaboration~\cite{JET:2013cls} and the previous JETSCAPE analysis based on inclusive hadron $\Raa{}$ only~\cite{JETSCAPE:2021ehl} (not plotted).

To isolate the impact of specific measurements on extraction of $\qhat{}$, the analysis is repeated for different classes of measurement. 
Figure~\ref{fig:hadronJetRaaPosteriors}, right panel, shows the result of such studies, presented as projections of the full $\qhat{}$ posterior distribution at fixed temperature and parton energy.
Posterior distributions are shown for jet $\Raa{}$ (blue), for charged hadron $\Raa{}$ (orange), and for charged hadron $\Raa{}$ with kinematic selection $\pT{}$ > 30 (green) and $\pT{} \leq$ 30 (red) \GeVc{}.
This figure shows two peaked distributions, where the $\pT{}$-inclusive hadrons and the low-$\pT{}$ hadrons consistently prefer a higher $\qhat{}$, while jets and high-$\pT{}$ hadrons prefer a smaller value.
While there is some overlap between the distributions, the apparent tension in most-probable value is due to the small experimental uncertainties of the low-$\pT{}$ hadron $\Raa{}$, which dominate the Bayesian inference analyses.
This tension points to the importance of including theoretical uncertainties in future studies and accounting for elements which are not yet included in the calculations, such as nuclear shadowing.


\section{Bayesian inference with inclusive jet $\Raa{}$ and jet substructure}

Expanding beyond the hadron and jet $\Raa{}$ study, we wanted to quantify the information gain from including jet substructure observables.
For this analysis, the $\qhat{}$ posterior is extracted from only the inclusive jet $\Raa{}$, and is compared to the posterior distribution obtained when including the jet $\Raa{}$, hadron-in-jet fragmentation $D(z)$~\cite{ATLAS:2017nre, CMS:2014jjt,ATLAS:2018bvp}, and groomed substructure observables $z_{\text{g}}$ and $R_{\text{g}}$~\cite{ALargeIonColliderExperiment:2021mqf}.
Since this is a proof-of-concept study, we including only published and unfolded observables in the most central event activity class, which also included a measured $\pp{}$ baseline.
For simplicity, correlation between some of the experimental uncertainties were neglected.

\begin{figure}[t]
    \centering
    \begin{tikzpicture}
        \node[anchor=south west, inner sep=0] (image) at (0,0) {\includegraphics[height=5.5cm]{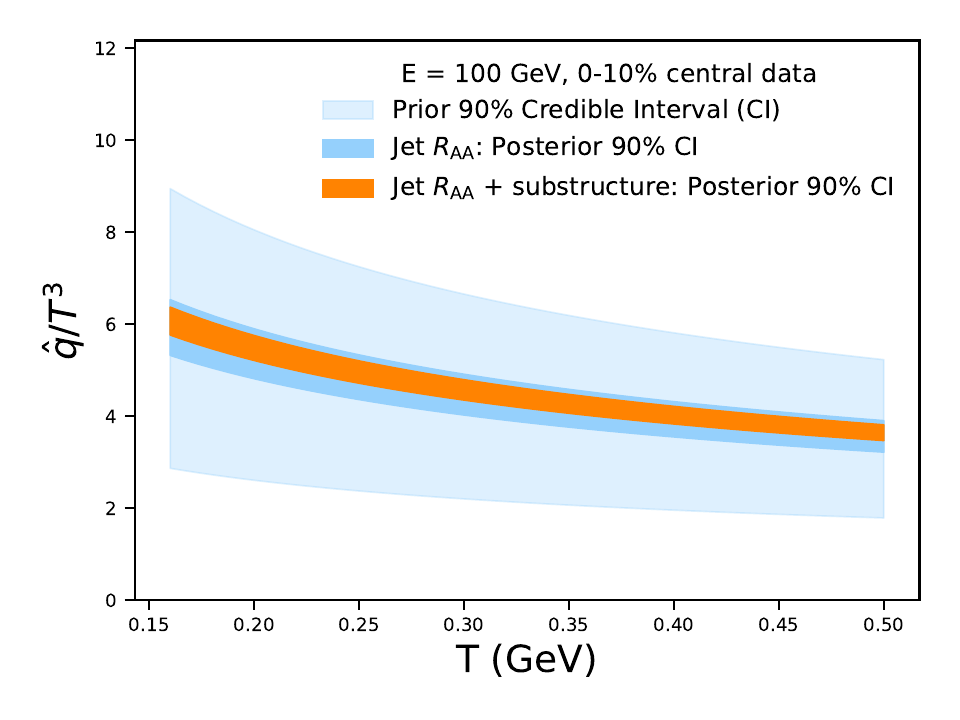}};
        \begin{scope}[x={(image.south east)},y={(image.north west)}]
            \node[anchor=south west, font=\sffamily\footnotesize] at (0.18, 0.2) {JETSCAPE Preliminary};
        \end{scope}
    \end{tikzpicture}
    \caption{Temperature dependent posterior distribution of $\qhat{}$ for only inclusive jet $\Raa{}$ (blue) and inclusive jet $\Raa{}$ and jet substructure (orange) using measurements performed for 0--10\% central collisions at $\sqrts{}$ = 200 GeV, 2.76 TeV, and 5.02 TeV.}
    \label{fig:jetSubstructure}
\end{figure}

The results from this study are shown in Fig~\ref{fig:jetSubstructure}.
The 90\% confidence interval of the posterior distribution extracted from jet $\Raa{}$ is shown in blue, while the posterior including the substructure observables is shown in orange.
Both cases show significant constraints with respect to the prior distribution.
While full conclusions will require expanding to include additional applicable measurements and centralities, the preliminary results show that including the substructure observables constrains $\qhat{}$ more strongly than only using the jet $\Raa{}$.
This indicates that there is additional information regarding jet quenching contained in the substructure measurements.
Note that the hadron-in-jet fragmentation observables include low-$\pT{}$ hadrons which introduced tension in the hadron and jet $\Raa{}$ measurement.
However, in this analysis, the extracted $\qhat{}$ posterior distributions are fully consistent despite including a subset of those hadrons.
Future studies will further characterize and disentangle the impact of hadron and jet observables on the extracted parameters.

\section{Summary}

We presented the results from two related Bayesian inference analyses utilizing inclusive hadron and jet observables to constrain the jet transport coefficient $\qhat{}$.
The first analysis shows that a consistent posterior distribution can be extracted for inclusive hadron and jet $\Raa{}$.
Further investigation indicates that small experimental uncertainties for low-$\pT{}$ hadron measurements dominate these calibrations.
Some apparent tension between the most probable $\qhat{}$ values preferred by low-$\pT{}$ hadrons and high-$\pT{}$ hadrons indicate missing uncertainties in the theoretical model.
This demonstrates the ability of these Bayesian techniques to provide feedback for models, as well the importance of including such uncertainties in future analyses.
A proof-of-concept study of the impact of including jet substructure observables indicates that these can provide more stringent constraints on $\qhat{}$, suggesting that these observables carry additional information about jet quenching.
Future analyses will continue to investigate further hadron and jet observables and additional jet quenching models, pinpointing regions of interest and providing feedback on model calculations.

\vspace{-0.2cm}

\section*{Acknowledgments}

\vspace{-0.1cm}

This work was supported in part by the National Science Foundation (NSF) within the framework of the JETSCAPE collaboration, under grant number OAC-2004571.

\vspace{-0.1cm}

%
\bibliography{rehlers_QM2023.bib}

\end{document}